\documentclass[preprint,showpacs,preprintnumbers,amsmath,amssymb]{revtex4}

\usepackage{graphicx}
\usepackage{dcolumn}
\usepackage{bm}

\begin{document}

\preprint{APS/123-QED}

\title{Pareto law of the expenditure of a person in convenience stores}

\author{Takayuki Mizuno$^{a}$, Masahiro Toriyama$^{b}$, Takao Terano$^{b}$, and Misako Takayasu$^{b}$}%
\address{$^{a}$Institute of Economic Research, Hitotsubashi University, 2-1 Naka, Kunitachi City, Tokyo 186-8603, Japan \\
$^{b}$Department of Computational Intelligence {\&} Systems Science, 
Interdisciplinary Graduate School of Science {\&} Engineering, 
Tokyo Institute of Technology, 4259 Nagatsuta-cho, Midori-ku, Yokohama 226-8502, Japan}

\date{\today}

\begin{abstract}
We study the statistical laws of the expenditure of a person in convenience 
stores by analysing around 100 million receipts. The density function of 
expenditure exhibits a fat tail that follows a power law. We observe the 
Pareto principle where both the top 25{\%} and 2{\%} of the customers 
account for 80{\%} and 25{\%} of the store's sales, respectively. Using the 
Lorenz curve, the Gini coefficient is estimated to be 0.70; this implies a 
strong economic inequality.\\


\end{abstract}


\maketitle

\section{Introduction}

 For the last ten years, high-frequency databases of market prices and 
company's wealth have been published. Physicists have investigated these 
databases in order to understand the dynamics of economy. They observed 
characteristics of fractal systems and chaotic behaviour in economic 
activities. For example, the probability density functions of market price 
fluctuations and company's incomes both follow power laws [1--5]. 

Recently, huge point-of-sale (POS) databases containing detailed records of 
each purchase were published. The analysis of POS databases revealed that a 
probability density function of the expenditure of a person in a single 
shopping trip also follows a power law [6].

In this paper, we confirm that a density distribution function of the 
expenditure of a person in a single shopping trip follows a power law with 
an exponent of --2 by analyzing a huge POS dataset of a convenience store 
chain. The frequency of trips that the person makes to the store chain 
depends on the individual. In order to clarify the extent of dependence of 
the store's sales on loyal customers, we analyze the expenditure of a person 
for three months. A density function of the expenditure exhibits a fat tail 
that follows a power law with an exponent -3. We observed that 80{\%} of the 
store's sales were accounted by approximately 20{\%} of the customers. 
Often, such a 80/20 rule has been observed in economic phenomena; this was 
referred to as the Pareto principle. The Pareto principle is important in 
business strategies. We estimate the Gini coefficient by plotting the Lorenz 
curve for the expenditure of a person for three months and discuss the 
dependence of a store's sales on loyal customers.

\section{Database}

The convenience store is a small shop that typically sells drinks, 
magazines, and prepackaged foods such as rice balls and lunchboxes. Also, 
many stores sell toiletries and cigarettes. Often, they stock daily 
necessities and miscellaneous goods. 

In this study, we analyzed a POS database in which a shopper's history in a 
convenience store chain was recorded. The name of this chain is ``am/pm 
Japan Co. Ltd.'' This chain has around 1,000 stores in Japan. The receipt 
number, date, time, the Japan article number (JAN), Edy ID, etc. were 
recorded in the database. Every JAN identifies a particular product. The Edy 
card provided by bitWallet, Inc. in Japan is a prepaid rechargeable card. 
Every Edy card has a unique number in the following format---Edy ID. When a 
payment is made using the Edy card, the Edy ID gets recorded in the POS 
database that is stored on a server. Each customer's history can be accessed 
through the recorded Edy ID; therefore, each customer's contribution to the 
store's profit can be ascertained.

We analyzed around 100 million receipts of all the customers of ``am/pm 
Japan Co. Ltd'' from January to March 2007. Out of all the receipts, 5{\%} 
of them are records corresponding to shopping done using the Edy card.

\section{Expenditure of a person in a single shopping trip}

Many people often buy lunch at a convenience store. Therefore, the average 
customer buys two food packets and one drink and spends around 400 yens. 
However, we can occasionally find high spending shoppers. For example, a 
young woman visited a store on December 28, 2006 at 9:28:35 AM, and bought 
1,127 drinks and 2,254 sandwiches at 559,590 yens. 

We show the distribution of the expenditure of a person in a single shopping 
trip by using about 100 million receipts in Fig. 1. The vertical axis, P ($ 
\ge $x), shows the cumulative probability in log scale, i.e., the 
probability of finding a person who spends more than x ¥ in a single 
shopping trip. It is found that the distribution has a fat tail that 
exhibits power law behaviour with an exponent of --2. The exponent is 
independent of the location of the store, the shopper's age, and the time of 
the day [6].

\section{Dependence of the store's sales on loyal customers}

 In order to investigate the extent of dependence of the store's sales on 
loyal customers, we analyzed the expenditure of a person for three months. 
We focussed on customers who paid using the Edy card. The expenditure of 
each customer for the three months can be calculated by searching the Edy ID 
in the POS database. Fig. 2 shows a cumulative density function of 
expenditure. This function can be approximated by a power law with an 
exponent -3 on the amount scale that is greater than $2 \times 10^{4}$ yens. On the 
small amount scale the function follows a lognormal distribution. 

 The Lorenz curve is one of the measures of social inequality in economics. 
We introduce the Lorenz curve that shows the bottom $x \times 100${\%} of the 
customers and the corresponding $y \times 100${\%} of the store's sales. If every 
customer spends the same amount of money, the bottom N{\%} of the customers 
would contribute to N{\%} of the sales. In that case, the Lorenz curve is 
depicted by a straight line y = x. However, if only one customer purchased 
something from the store, the curve depicts a line of perfect inequality 
given by $y = 0$ for all $x < 1$, and $y = 1$ when $x = 1$. 

Fig. 3 depicts the Lorenz curve of the customer's payment and the store's 
sales for three months. The curve is strongly convex downward implying that 
the store's sales strongly depend on the loyal customers. We observe that 
both the top 25{\%} and 2{\%} of the customers account for 80{\%} and 25{\%} 
of the sales, respectively. This 80/20 rule is referred to as the Pareto 
principle. The original Pareto principle was related to wealth. It is well 
known that 80{\%} of a country's wealth is owned by 20{\%} of the 
population. If the expenditure of a person in the convenience store depends 
on the person's wealth, it is evident that the expenditure follows the 80/20 
rule.

The Gini coefficient is often used as a measure of inequality of wealth 
distribution. Twice the area between the line of perfect equality, y = x, 
and the observed Lorenz curve gives an estimate of the Gini coefficient. The 
coefficient is zero when everyone's expenditure is the same, and is one when 
only one person purchases everything from the convenience store chain. The 
Gini coefficient estimated from the Lorenz curve in Fig. 3 is 0.70. It is 
usually observed that the coefficient is less than 0.4 in a market economy. 
Therefore, the coefficient of 0.70 means that there is significant 
inequality in the store's sales. From the coefficient, we can also find that 
the loyal customers contribute significantly to the store's sales.

\section{Discussion}

 In this paper, we analyzed a huge point-of-sale database that has detailed 
records of each purchase. This data is the latest target of econophysics, 
following the ``market price'' and ``company's wealth.'' It is confirmed 
that a density function of the expenditure of a person in a single shopping 
trip follows a power law with an exponent of --2. Such a power law is a 
commonly observed characteristic with regard to the market price and a 
company's wealth.

We analyzed the expenditure of a person for three months. A cumulative 
density function of the expenditure exhibits a fat tail that follows a power 
law with an exponent -3. We investigated the extent of the dependence of the 
store's sales on loyal customers by using the Lorenz curve and the Gini 
coefficient. We found that both the top 25{\%} and 2{\%} of the customers 
account for 80{\%} and 25{\%} of the store's sales, respectively. A 
marketing strategy aimed at high-spending shoppers may be effective for 
convenience store chains because they contribute significantly to a store's 
sales. 

The statistical laws mentioned in this paper are often observed in various 
other economic phenomena. Analysis of a POS database may be useful for 
investigating the universal dynamics of the economic phenomena.

\begin{acknowledgments}
This work is partly supported by Research Fellowships of the Japan Society 
for the Promotion of Science for Young Scientists (T. Mizuno.). The authors 
appreciate the cooperation of ``am/pm Japan Co. Ltd'' for providing the POS 
data.
\end{acknowledgments}

\newpage

\begin{figure}
\vspace{6cm}

\centerline{\includegraphics[width=3.89in,height=3.26in]{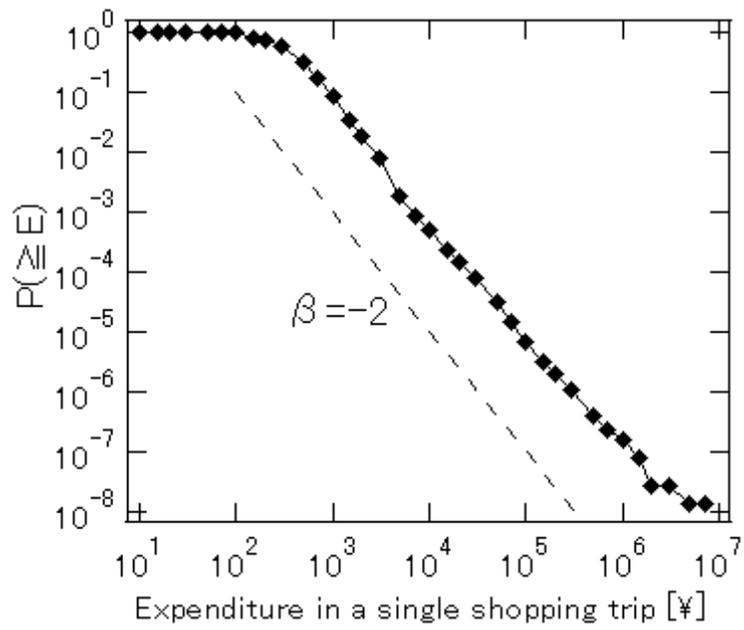}}
\caption{Cumulative density function of the expenditure of a person in a 
single shopping trip. The guideline follows a power law with an exponent of 
--2.}
\end{figure}

\begin{figure}

\vspace{6cm}
\centerline{\includegraphics[width=4.41in,height=3.60in]{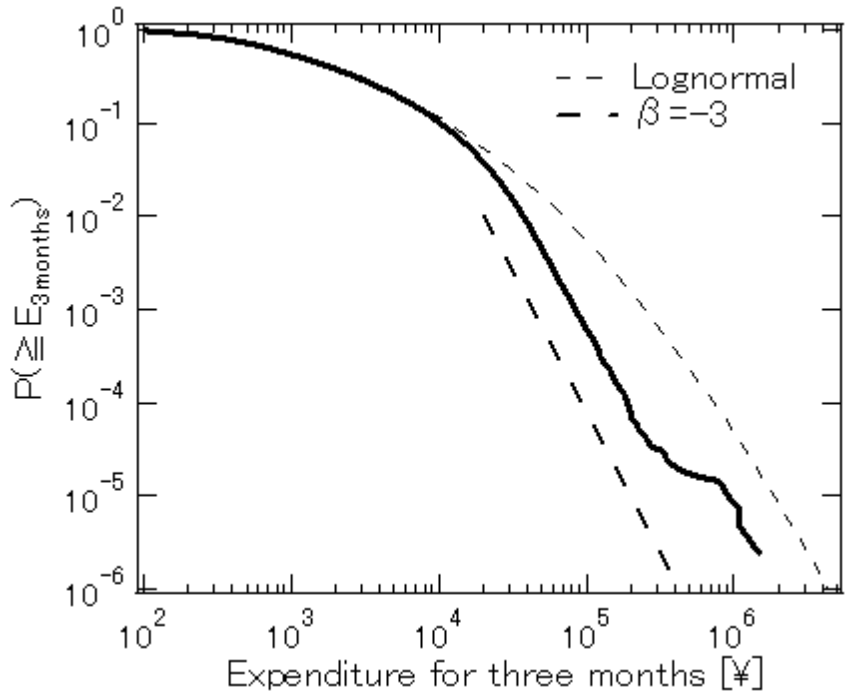}}
\caption{Cumulative density function of the expenditure of a person for three 
months. Here, we have depicted only the payments made using the Edy card. 
The straight guideline expresses a power law with an exponent of --3. The 
curved guideline indicates a lognormal distribution with $<$log$ E_{3months}>=3.13$ and 
$\sigma ($log$ E_{3months})=0.74$. log$E_{3months}$ indicates the logarithmic expenditure for three months.}
\end{figure}

\newpage

\begin{figure}
\centerline{\includegraphics[width=4.65in,height=3.66in]{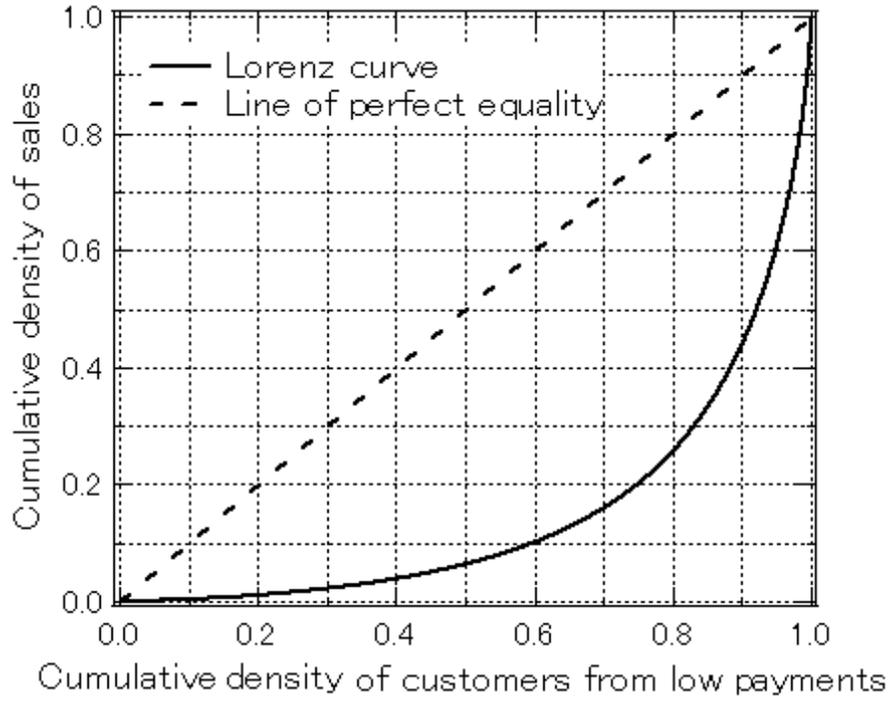}}
\caption{The Lorenz curve of a customer's payment and a store's sales for 
three months. The Lorenz curve and the line of perfect equality are shown in 
the figure.}
\end{figure}

\end{document}